\documentclass[twocolumn]{aastex631}
\shorttitle{Modeling the high-brightness state of the recurrent nova T~CrB}
\shortauthors{Schlindwein, Baptista \& Luna}
\graphicspath{{./}{figures/}}
\usepackage{soulutf8}

\newcommand{\xmm}{\textit{XMM}-Newton}

\newcommand{\tcrb}{T~CrB}

\begin{document}

\title{Modeling the high-brightness state of the recurrent nova T~CrB as an enhanced mass-transfer event}

\correspondingauthor{Wagner Schlindwein}
\email{wagner.schlindwein@astro.ufsc.br}

\author[0000-0002-7095-4147]{Wagner Schlindwein}
\affiliation{Instituto Nacional de Pesquisas Espaciais, Avenida dos Astronautas, 1758, São José dos Campos-SP, Brazil}
\affiliation{Departamento de F\'isica, Universidade Estadual de Maring\'a, Av. Colombo, 5790, Maring\'a-PR, Brazil}

\author[0000-0001-5755-7000]{Raymundo Baptista}
\affiliation{Departamento de F\'isica, Universidade Federal de Santa Catarina, Campus Trindade, Florianópolis-SC, Brazil}

\author[0000-0002-2647-4373]{Gerardo Juan Manuel Luna}
\affiliation{Universidad Nacional de Hurlingham (UNAHUR). Secretaría de Investigación, Av. Gdor. Vergara 2222, Villa Tesei, Buenos Aires, Argentina}
\affiliation{Consejo Nacional de Investigaciones Científicas y Técnicas (CONICET).}



\begin{abstract}

T~Coronae Borealis is the nearest symbiotic recurrent nova. Twice in the last two centuries, in 1866 and 1946, the accreted material ignited on the surface of the white dwarf via runaway thermonuclear fusion reactions and produced a nova eruption. 
Both eruptions occurred approximately midway through a transient state of high luminosity. A possible explanation of such a state is a dwarf-nova-like outburst, which may arise from a transient increase in the mass-transfer rate of the donor star. 
We simulate the response of an accretion disk to an event of enhanced mass-transfer that is ``interrupted'' by a pre-eruption dip associated to the convective phase leading to the thermonuclear runaway, and model the resulting optical light curve using the parameters of the T~CrB binary. Our model represents the first attempt to reproduce the transient high-accretion state.
The observed brightening can be satisfactorily reproduced by models of an accretion disk with a viscosity parameter $\alpha = 3$, an event of enhanced mass-transfer with a duration of $\Delta t = 15$\,yr, and quiescent and high-state mass-transfer rates of $2.0 \times 10^{-9} \, M_\odot$\,yr$^{-1}$ and $1.9 \times 10^{-7} \, M_\odot$\,yr$^{-1}$, respectively, while the pre-eruption dip can be reproduced by the small, accelerated expansion of the inner disk radius, at an average velocity of 0.02\,km\,s$^{-1}$.
Our model is also capable of reproducing the observed changes in color of T~CrB throughout the transient event.

\end{abstract}

\keywords{Stellar accretion disks (1579) --- Symbiotic binary stars (1674) --- Recurrent novae (1366) --- Astronomical simulations (1857) --- Interacting binary stars (801)}


\section{Introduction} \label{sec:int}

T~Coronae Borealis (T~CrB) is the nearest symbiotic recurrent nova, at a distance of 916\,pc \citep{Gaia2016,GaiaDR3}. The system contains a M4~III giant \citep[RG,][]{Zamanovet2023}, which fills its Roche lobe and transfers mass to a massive white dwarf (WD) with $M_1 \geq 1.1 M_\odot$ \citep{1998MNRAS.296...77B,2018ApJ...860..110S,2019ApJS..242...18H}. This transferred mass forms an accretion disk around the WD with a size of $89 \pm 19 \, R_\odot$ \citep{Zamanovet2024}. Its two well-recorded outbursts in 1866 and 1946, together with historical eruptions in 1217 and 1787, support a recurrence time of roughly 80 years \citep{Schaefer2023}. During these occasions, the accreted material ignited on the surface of the white dwarf through runaway thermonuclear fusion reactions and produced a nova eruption.

Approximately eight years before the 1866 and 1946 eruptions, the optical brightness of \tcrb\ increased by about 1.5 magnitudes in the $B$-band and stayed at that level until about 8 months before the nova eruption. In fact, by studying the years before and after both previous thermonuclear runaways (TNRs), \citet{schaefer2010,Schaefer2023} found that \tcrb\ brightening started about 8 years {\em before} the eruption and lasted through about 7 years {\em after} the eruption, being interrupted for about 1 year {\em before} the eruption when the brightness returned to quiescent values (this phase has now been coined as the pre-eruption dip) and 1 year {\em after} the eruption until returning to quiescence \citep[see Figure~23 in][]{schaefer2010}. From late 2014 until December 2022, T~CrB was in a brightness akin of the similar brightening state that then started to fade and reached a brightness similar to its quiescence level in October 2023 \citep{munari23,Zamanovet2023}. Figure~\ref{fig:dados} shows the recent optical behavior of T~CrB, overlaid on the light curve from around the time of the 1946 eruption but shifted by -78 years. The alignment between both light curves is remarkable and most likely the current high-accretion state would lead to the next nova eruption, as did in the two previous eruptions. The comparison between current and previous brightness behavior has allowed different authors to predict that the next nova eruption will occur within 2025$\pm$2\,yr \citep{Lunaet2020,Schaefer2019,Zamanovet2023}.

There is mounting evidence that this increase in brightness is due to an increase in the accretion rate \citep{Lunaet2018,Schaefer2023,Ilkiewiczet2023}. UV observations during the low-accretion state allowed to derive accretion rates of $\dot{M}_\mathrm{low} = 2.0 \times 10^{-9} \, M_\odot$\,yr$^{-1}$ \citep{selvelli92,Stanishevet2004}. From the analysis of \xmm\ observations taken a few months after the start of this state, known as the superactive state after \citet{Munariet2016}, \citet{Lunaet2018} found a new super-soft component in the X-ray spectrum with a luminosity too low to be due to nuclear burning on the white dwarf surface, but consistent with the optically thick emission from the boundary layer of the accretion disk, and estimated a lower limit to the accretion rate during the high-brightness state of $8.5 \times 10^{-8} \, M_\odot$\,yr$^{-1}$. In addition, the spectroscopic analysis of the recent high-brightness state by \citet{Planquartet2025} demonstrates that this state correlates with the bright spot enhancement, and that the increase in bright spot luminosity must originate from either a decrease in the outer disk radius or an increase in the mass-transfer rate.

Given the recurrent nova nature of \tcrb\, the ignition of the following nova eruption requires the previous accumulation of a hydrogen-rich envelope of accreted matter, $M_{ig}$, onto its WD surface \citep[e.g.,][]{Fujimoto1982,ShenBildsten2009,Wolfetal2013}. $M_{ig}$ depends on $M_1$ and on the WD mass accretion rate \citep[as well as on chemical composition, see][]{ShenBildsten2009}. A larger WD mass leads to higher accreted envelope temperatures and densities, which allows for the ignition conditions to be reached at lower envelope masses, whereas higher accretion rates lead to higher compressional heating, also resulting in ignition occurring at lower envelope masses.

In the case of \tcrb\, the envelope mass required to trigger a nova eruption onto a WD of mass $M_1=1.3\,M_\odot$ fed by a solar composition gas at a quiescent accretion rate of $2\times 10^{-9} \, M_\odot$\,yr$^{-1}$ is $M_{ig}\simeq 1.1\times 10^{-5} \, M_\odot$ \citep{ShenBildsten2009}. The timescale required to accumulate this envelope mass at the quiescent accretion rate is $\simeq 5480$\,yr, almost two orders of magnitude longer than the observed $\approx$80\,yr recurrence time. This implies that {\em there must be a phase of significantly enhanced mass accretion} which reduces $M_{ig}$ and shortens the timescale required to trigger the nova eruption down to the observed recurrence time. Indeed, at a higher mass accretion rate of $\simeq 1.7 \times 10^{-7} \, M_\odot$\,yr$^{-1}$ the required envelope mass is reduced to $M_{ig}\simeq 2.6\times 10^{-6} \, M_\odot$ and the ignition conditions can be reached after only $\simeq 15$\,yr, which is in good agreement with the full length of the high-brightness state. This reasoning suggests that a high-mass accretion phase is key to explain the timings and recurrent nova behavior of T~CrB, and sets the motivation for this paper.

Here, we model the observed high-brightness state as an event of enhanced mass-transfer rate from the RG, with the additional requirement that the matter accreted over a time length of $(80\pm 2)$\,yr equals the required envelope mass needed to trigger nova eruptions at the observed recurrence interval. The data is described in Section~\ref{sec:obs}; the details of the model and its results are presented in Section~\ref{sec:res}. The results are discussed in Section~\ref{sec:disc} and a summary of our conclusions is presented in Section~\ref{sec:concl}.

\section{Observations} \label{sec:obs}

\begin{figure}[!ht]
\includegraphics[width=1.0\columnwidth]{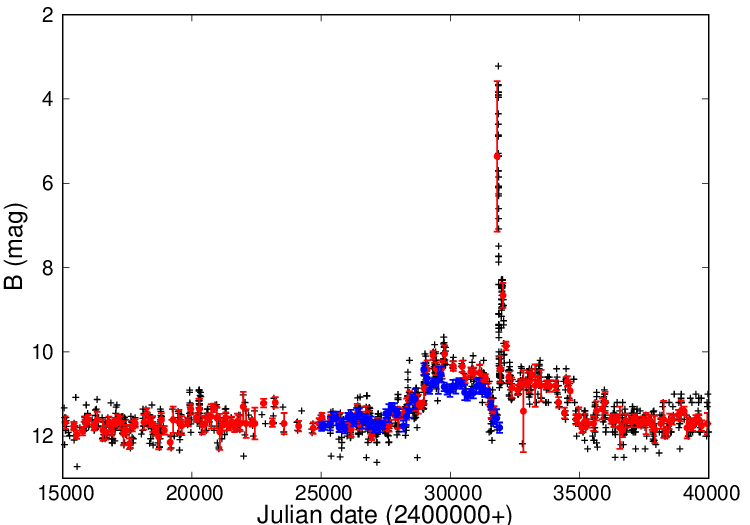}
\caption{$B$-magnitudes light curves of T~CrB during the 1938-1955 brightening state including the 1946 thermonuclear runaway (black crosses). We include the AAVSO data from the 2004 until April 2024 -- brightening state shifted by -78\,yr and with a 113-days median filter (blue dots with error bars). The same filter is used for the historical data (red dots with error bars).
\label{fig:dados}}
\end{figure}

In modeling, we used the historical dataset of T~CrB compiled by \citet{Schaefer2023_mnras} that cover the brightness state before and after the 1946 nova eruption, and $BVRI$-band observations from the archive of the American Association of Variable Star Observers (AAVSO) from 2004 until April 2024. In Figure~\ref{fig:dados} the historical light curve is shown as black crosses. In order to retain only the long-term trend, we smoothed the historical and AAVSO light curves by applying a median filter with a width of 113 days (half orbital period; red and blue points with error bars in Figure~\ref{fig:dados}, respectively). The historical data include both the brightening that started around 1938 and the nova eruption that occurred in 1946, while the AAVSO data include the current brightening event and were shifted by -78 years to match the start of the current brightening with that of the 1938 event.

Before the start of its recent high-accretion state, T~CrB had average low-accretion state magnitudes of $B = 11.67 \pm 0.10$\,mag, $V = 10.17 \pm 0.07$\,mag, $R = 9.04 \pm 0.10$\,mag and $I = 7.52 \pm 0.06$\,mag. It is worth noting that we consider the entire period of high-accretion before and after the nova eruption as a single transient, and that the brightness decrease prior to the eruption is caused by another phenomenon (see discussion in Section~\ref{sec:disc}).

\section{Model and results} \label{sec:res}

The 1946 nova eruption occurred during a transient high-accretion state (see Figure~\ref{fig:dados}). A natural explanation for this state is a dwarf-nova-like outburst, where a possible origin of the increase in the accretion rate is an instability in the mass-transfer rate from the donor star leading to an event of mass-transfer increase \citep[Mass-Transfer Instability Model, MTIM,][]{Bath1972,Bath1975,BathPringle1981}. 

MTIM assumes that dwarf nova outbursts are the response of a disk with constant (and high) viscosity to a sudden increase in the mass-transfer rate ($\dot{M}_2$). Therefore, to model the high-accretion state of T~CrB, a high-viscosity is required for a quick response of its accretion disk to the event of enhanced mass-transfer.

We simulate the response of the accretion disk to events of enhanced mass-transfer, in the context of the MTIM, to model the optical light curve of T~CrB. We assume the viscosity parameter $\alpha$\footnote{The solution is known as the steady $\alpha$-disk model and is obtained by adopting the prescription of \citet{ShakuraSunyaev1973}, $\nu = \alpha c_s H$, where $\alpha$ is a dimensionless parameter, $c_s$ is the speed of sound, and $H$ is the disk vertical scale height.} to be constant, with a high value ($\geq 1$). The reader is referred to  \citet{SchlindweinBaptista2024} for further details on the calculation of the time evolution of the accretion disk, and to \citet{BaptistaSchlindwein2022} and \citet{SchlindweinBaptista2024} for a discussion on the plausibility of $\alpha \geq 1$ and its implications. Since we are interested in modeling only the luminosity of the disk, we assume that matter is deposited at its outer edge, uniformly over a radial extent of $\Delta R= 0.1\,R_d$, where $R_d$ is the outer disk radius. We adopted $\mu = 0.615$ (suitable for a solar abundance of fully ionized gases) and a tidal truncation constant of $c\omega= 0.01$\,rad/s \citep[Equation~5 from][]{IchikawaOsaki1992} which corresponds to a quiescent disk radius of $R_d \sim 0.74\,R_{L_1}$ (where $R_{L_1}$ is the distance from the WD to the inner Lagrangian point $L_1$), consistent with the value inferred by \citet{Zamanovet2024}. As discussed in Section~2.2 of \citet{SchlindweinBaptista2024}, since the brightness level during the high-accretion state presents a plateau during maximum (Figure~\ref{fig:dados}), it is appropriate for us to adopt an enhanced mass event with the shape described by:

\begin{equation}
\dot{M}_2 = \dot{M}_2^q + (\dot{M}_2^h-\dot{M}_2^q) \exp \left[ -\frac{1}{2} \left( \frac{t-t_e}{b \, \Delta t_e} \right)^n \right],
\label{eq:pulso}
\end{equation}

\noindent where $\dot{M}_2^q$, $\dot{M}_2^h$, $t_e$ and $\Delta t_e$ are the quiescent and high-accretion state mass-transfer rates, the event center instant, and its full width at half maximum (FWHM), respectively. The addition of the parameter $n$ (which must be even) in the exponent of the exponential argument is used to adjust the plateau of the enhanced mass-transfer event and, as a consequence of this parameter, a term $b=[2(2 \ln 2)^{1/n}]^{-1}$ was added to ensure that $\Delta t_e$ is the FWHM. We considered that the disk emits locally as a blackbody. The resulting spectra were convolved with the $BVRI$ passbands responses to derive the corresponding magnitudes.

As the RG contributes significantly to the total luminosity of the system, a good estimate of its magnitudes and variations becomes important for the results of the model. For the $V$ passband, the linear term of the ellipsoidal variability of \citet{Zamanovet2023} is $V_\mathrm{RG} = 10.16$\,mag. For the other passbands, we assume an M4~III spectral type for the RG \citep{Zamanovet2023} and adopt color indexes of $B-V = 1.55$\,mag, $V-R = 1.17$\,mag and $V-I = 2.69$ \,mag \citep{Allerbook1996}. The RG is thus by far the dominant light source in the $V$, $R$ and $I$ passbands during the low-accretion state, with $B_\mathrm{RG}= 11.71$\,mag, $R_\mathrm{RG}= 8.99$\,mag and $I_\mathrm{RG}= 7.47$\,mag. In order to take into account the irradiation effect of the accretion luminosity onto the RG, we use the irradiation model of \citet{Hameuryet2020}, adopting the albedo $\eta = 0.6$. Their model requires the assumption of an (irradiation free) effective temperature $T_2$ for the irradiated star. However, the spectrum of the RG is far from that of a blackbody and it is not possible to represent its $BVRI$ magnitudes with a single effective temperature. Therefore, for each passband, we adopted a $T_2$ value which matches the observed RG brightness at that passband (see Table~\ref{tab:parametros_binarios}).

\begin{deluxetable*}{ccc}
\tablenum{1}
\tablecaption{T~CrB adopted binary parameters and fitted model parameters. \label{tab:parametros_binarios}}
\tablewidth{0pt}
\tablehead{
Parameters & Model~I & Model~II
}
\decimalcolnumbers
\startdata
Orbital period & \multicolumn{2}{c}{$5461.65$\,h $^\mathrm{a}$} \\
$M_1$ & $1.37 \, M_\odot$ $^\mathrm{b}$ & $1.29 \, M_\odot$ \\
$R_1$ & $0.003 \, R_\odot$ & $0.0045 \, R_\odot$ \\
$M_2$ & $1.12 \, M_\odot$ $^\mathrm{b}$ & $0.70 \, M_\odot$ \\
$R_2$ & $76.5 \, R_\odot$ & $64.3 \, R_\odot$ \\
Inclination ($i$) & $67^\circ$ $^\mathrm{b}$ & $57.3^\circ$ \\
Mass ratio ($q$) & 0.82 $^\mathrm{b}$ & 0.54 \\
Binary separation ($a$) & $211.5 \, R_\odot$ & $196.2 \, R_\odot$ \\
$R_{L_1}$ & $110.0 \, R_\odot$ & $110.1 \, R_\odot$ \\ 
$\mu$ & \multicolumn{2}{c}{0.615} \\
Distance & \multicolumn{2}{c}{916\,pc $^\mathrm{c}$} \\
$T_2^B$ & 2967\,K & 3065\,K \\
$T_2^V$ & 2910\,K & 3029\,K \\
$T_2^R$ & 2914\,K & 3056\,K \\
$T_2^I$ & 3080\,K & 3277\,K \\
$\eta$ & \multicolumn{2}{c}{0.6} \\
$\dot{M}_2$ in quiescence ($\dot{M}_2^q$) & \multicolumn{2}{c}{$2.0 \times 10^{-9} \, M_\odot$\,yr$^{-1}$ $^\mathrm{d}$} \\
$\dot{M}_2$ during event maximum ($\dot{M}_2^h$) & $1.7 \times 10^{-7} \, M_\odot$\,yr$^{-1}$ & $1.9 \times 10^{-7} \, M_\odot$\,yr$^{-1}$ \\
Viscosity parameter ($\alpha$) & \multicolumn{2}{c}{3.0} \\
Event duration ($\Delta t_e$) & \multicolumn{2}{c}{15\,yr} \\
Plateau parameter ($n$) & \multicolumn{2}{c}{8} \\
\enddata
\tablecomments{a - \citet{Fekelet2000}; b - \citet{Stanishevet2004}; c - \citet{GaiaDR3}; d - \citet{selvelli92,Stanishevet2004}.}
\end{deluxetable*}

\citet{Munariet2016} used the T~CrB infrared light curves to infer a temperature increase of the RG heated side at the high-brightness state of $\sim 80$\,K, or about a 2\% increase in comparison to the unirradiated side, implying an irradiating flux of $\sim 8$\% of the intrinsic stellar flux. Using the \citet{Hameuryet2020} model to compute the irradiation effect, we find that irradiation by the accretion luminosity increases the surface temperature of the RG heated side by $\sim 1$\,K and $\sim 100$\,K in quiescence and in the high-brightness state, respectively, resulting in an RG $\sim 0.10$\,mag brighter in the $I$ passband during the high-brightness state -- in good agreement with the brightness increase inferred by \citet{Munariet2016}.

The parameters adopted for our model~I simulation are listed in Table~\ref{tab:parametros_binarios}. The contribution of the WD emission to the $BVRI$ passbands is considered negligible. Interstellar extinction effects were included; we assumed the extinction law of \citet{Cardelliet1989} and adopted $E(B-V) = 0.056 \pm 0.003$\,mag/kpc \citep{SchlaflyFinkbeiner2011}. The $t_e$ and $\Delta t_e$ values are inferred directly from the historical light curve (Figure~\ref{fig:dados}), while the values of $\dot{M}^q_2$ and $T_2$ are obtained by fitting the low-accretion $BVRI$ brightness level. Accordingly, we confirm that the $\dot{M}_\mathrm{low}$ value of \citet{Lunaet2018} (derived from a mean between values from \citet{selvelli92} and \citet{Stanishevet2004}) provides a good description of the T~CrB quiescent brightness level. The remaining model parameters were fitted through a set of reduced simulations optimized for this purpose. $\alpha \geq 1$ are required in order to match the observed transient decline timescale in T~CrB, whereas for $\alpha > 3$ the differences in the decline rate are small compared to the uncertainties of the data.

\begin{figure}[!ht]
\centering
\includegraphics[width=1.0\columnwidth]{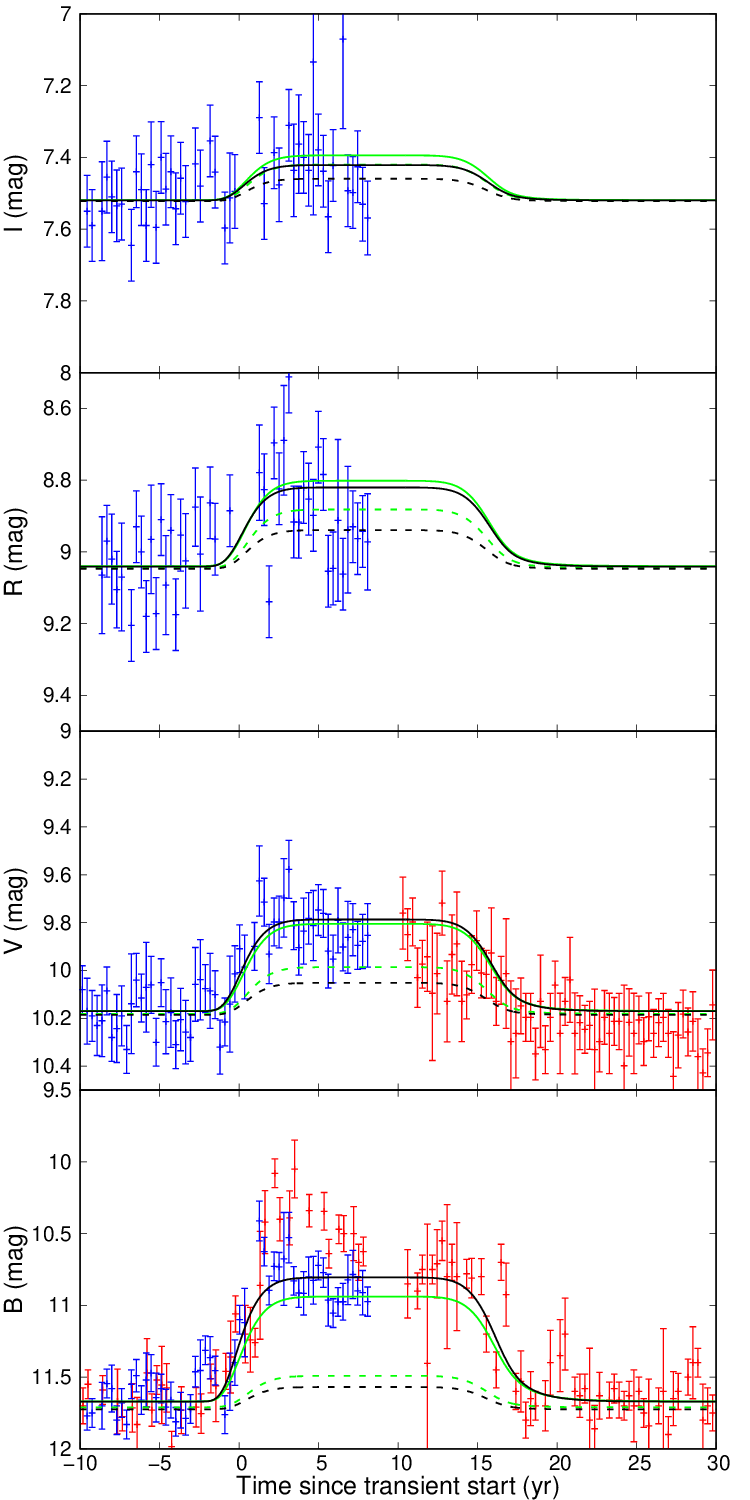}
\caption{Simulation results with model~I (green solid curve) and model~II (black solid curve) parameters. The points with error bars are the same as in Figure~\ref{fig:dados} for $B$ band and analogous for the other bands. The green and black dashed curves are the contribution from the irradiated red giant companion to model~I and II, respectively. \label{fig:simulacao}}
\end{figure}

The best-fit $BVRI$ light curves corresponding to model~I are shown in Figure~\ref{fig:simulacao}. Note that these light curves are the sum of the brightness of the simulated accretion disk plus the contribution from the irradiated RG companion. The high-accretion state is dominated directly by the accretion disk outburst in the $B$ and $V$ passbands, but by the resulting RG irradiation effects in the $R$ and $I$ passbands. The inferred accretion rate during the high-accretion state is $1.7 \times 10^{-7} \, M_\odot$\,yr$^{-1}$, twice the lower limit derived by \citet{Lunaet2018}.

For a WD of mass $1.37 \, M_\odot$, the mass of the accreted envelope required for a nova eruption ignition is $M_{ig} \simeq 10^{-6} \, M_\odot$ for accretion rates $\simeq 7 \times 10^{-8} \, M_\odot$\,yr$^{-1}$ \citep{ShenBildsten2009}. Considering that the recurrence time of the nova eruption in T~CrB is $\sim 80$\,yr, that for about 15\,yr the system is in a high-accretion rate state, and that for the remainder of the time it is in a quiescent state accreting at a rate of $2.0 \times 10^{-9} \, M_\odot$\,yr$^{-1}$, the accretion rate required at the high-accretion state is $\simeq 5.8 \times 10^{-8} \, M_\odot$\,yr$^{-1}$. This upper limit is $\sim 3 \times$ smaller than inferred by model~I and would result in the system being $\sim 0.2$\,mag fainter than model~I in the $V$ passband during the high-accretion state, with the accretion disk and the RG being, respectively, $\sim 0.75$\,mag and $\sim 0.13$ fainter than in model~I. The WD mass in T~CrB is quite uncertain, and this discrepancy suggests that the WD mass choice for model~I might be overestimated, prompting us to obtain additional constraints on $M_1$.

\begin{figure}[!ht]
\centering
\includegraphics[width=1.0\columnwidth]{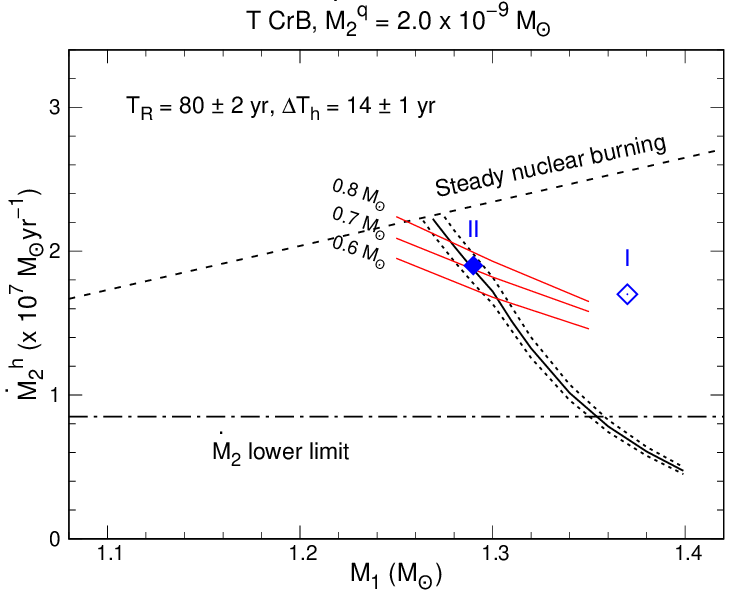}
\caption{Diagram of accretion rate versus WD mass for T~CrB. The upper black dashed line represents the limit above which steady nuclear burning occurs. The lower limit in $\dot{M}$ \citep[from][]{Lunaet2018} is indicated by the black dashed-dotted line. The black solid and dotted lines mark the relationship $\dot{M} (M_1)$ from the envelope mass that produces a nova eruption, where $T_R = 80 \pm 2$\,yr is the assumed value of the nova eruption recurrence interval, and $\Delta T_h = 14 \pm 1$\,yr is the time spent in the high-accretion state. The red solid lines indicate the relationship $\dot{M} (M_1)$ adopting $M_2 = 0.6$, 0.7 and 0.8\,$M_\odot$ in the T~CrB primary mass function of \citet{Fekelet2000}. Open and filled blue diamonds mark the solutions for models I and II, respectively.\label{fig:diagrama}}
\end{figure}

A diagram of the high-accretion state mass-transfer rate $\dot{M}_2^h$ required to reach the corresponding $M_{ig}$ value at the end of the recurrence interval is plotted as a function of WD mass in Figure~\ref{fig:diagrama}. The relationship $\dot{M} (M_1)$ was estimated using the curves from \citet{ShenBildsten2009}, assuming a value of $80 \pm 2$\,yr for the nova eruption recurrence interval, a time spent in the high-accretion rate state of $14 \pm 1$\,yr, and the quiescence accretion rate of $2.0 \times 10^{-9} \, M_\odot$\,yr$^{-1}$ \citep{Lunaet2018}. The limit above which steady nuclear burning occurs, i.e., there no longer are recurrent nova eruptions, comes from the expression $\dot{M} (\textrm{crit}) \approx 2.3 \times 10^{-7} (M_1/M_\odot - 0.19)^{3/2}\, M_\odot$\,yr$^{-1}$ \citep{Warner1995}. The lower limit for $\dot{M}$ comes from \citet{Lunaet2018}. Consistent model solutions are along the solid line; solutions below and to the left (above and to the right) of this line imply recurrence intervals longer (shorter) than 80\,yr. With this constraint in mind, the range of possibilities for $M_1$ becomes reasonably narrow, between 1.27 and 1.35\,$M_\odot$. Lower $M_1$ values put the system in the steady nuclear burning region with no recurrent nova events, while higher $M_1$ values lead to nova recurrence intervals shorter than observed. Thus, the $M_1=1.37\,M_\odot$ and $\dot{M}_2^h= 1.7 \times 10^{-7} \, M_\odot$\,yr$^{-1}$ solution of model~I is not self-consistent because it implies a recurrence interval shorter than observed.

Aside of the WD mass, the inclination is another poorly constrainted binary parameter which significantly influence the results of the T~CrB simulations. Tests were conducted with the intent of characterizing the effects of $M_1$ and $i$ in the T~CrB simulations. A smaller primary mass implies a fainter accretion disk while the irradiation effect decreases, as $T_\mathrm{2,irr}$ depends on $M_1^{0.25}$ and $R_1^{-0.25}$. On the other hand, a smaller inclination implies that the apparent disk brightness increases while the irradiation effect remains unchanged. Therefore, with a less massive WD, a higher mass-transfer rate is required to describe the transient, whereas with a smaller inclination, a lower rate is necessary.

We can use Figure~\ref{fig:diagrama} as a guide to estimate not only $M_1$ but also the inclination of T~CrB with a relatively small uncertainty range. To obtain the binary parameter set for this test, we adopted the T~CrB primary mass function of \citet{Fekelet2000}, where the other model parameters ($q$, $a$, $R_{L_1}$, $R_2$, $T_2$) were modified to be consistent with the new values of $M_1$ and $i$. With this constraint, the binary inclination must be in the range $45^\circ \lesssim i \lesssim 70^\circ$, as the system is non-eclipsing.

This exercise leads to model~II, with $M_1 = 1.29 \, M_\odot$, $i = 57.3^\circ$, and a best-fit mass accretion rate at the high-accretion state of $\dot{M}_2^h = 1.9 \times 10^{-7}\, M_\odot$\,yr$^{-1}$. The parameters adopted for the model~II simulation are listed in Table~\ref{tab:parametros_binarios}, and its position is marked as a blue filled diamond in Figure~\ref{fig:diagrama}. It results in a quiescent disk radius of $R_d \sim 0.70\,R_{L_1}$ \citep[in good agrement with the value inferred by][]{Zamanovet2024} while the irradiation by the accretion luminosity increases the surface temperature of the RG heated side by $\sim 0.5$\,K and $\sim 70$\,K in quiescence and in the high-accretion state, respectively, resulting in an RG $\sim 0.06$\,mag brighter in the $I$ passband during the high-accretion state. The major contribution to the high-accretion state in model~II comes from the accretion disk outburst even in the $I$ passband. Since in model~II the radius of the RG is considerably smaller, larger $T_2$ values are required in order to obtain the same observed flux.

The best-fit $BVRI$ light curves corresponding to model~II are shown in Figure~\ref{fig:simulacao}. Model~II provides a better description of the observations than model~I, while consistently falling onto the $\dot{M}(M_1)$ relationship of Figure~\ref{fig:diagrama}. Model~II is bluer than model~I. This is due to the fact that the amplitude of the RG irradiation effect is smaller in this model, thus requiring a more pronounced modulation of the brightness of the accretion disk.

\begin{figure}[!ht]
\includegraphics[width=1.0\columnwidth]{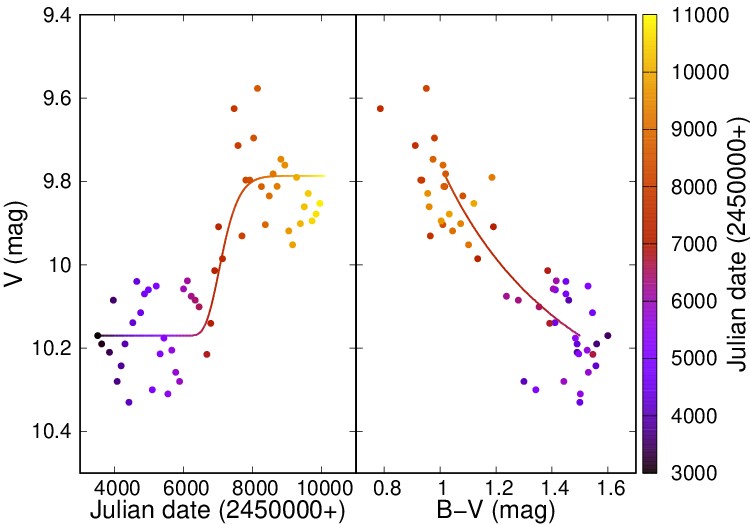}
\caption{\textit{Left-hand:} T~CrB light curve in the $V$ band. \textit{Right-hand:} Colour variations across the most recent high-accretion state of T~CrB. The dots are the AAVSO data and the solid line is the result of the simulations of model~II, while the colour-coding represents the time. \label{fig:color}}
\end{figure}

Figure~\ref{fig:color} shows the smoothed version of the AAVSO $V$-passband light curve and the corresponding colour-magnitude diagram. The statistical uncertainties of the $(B-V)$ colour are typically around 0.18\,mag. In quiescence, $(B-V) \simeq 1.5$; T~CrB becomes bluer during the high-accretion state, with a $(B-V) \simeq 1.0$. The predicted colour variations of model~II are shown as a solid line in the right-hand panel of Figure~\ref{fig:color}, and provide a good description of the observed color variations for the rise to the high-accretion state.

\citet{Schwarzenberg1981} shows that chromospheric emission is relevant for accretion rates $\dot{M} > 10^{-8} \, M_\odot$\,yr$^{-1}$ such as those inferred for the high-accretion state of T~CrB, with a predicted increase in $V$ passband brightness by $\sim 0.2$\,mag for an accretion disk with $\dot{M} = 10^{-7} \, M_\odot$\,yr$^{-1}$ and $i = 60^\circ$. While the inclusion of chromospheric emission is beyond the scope of the present paper, one might expect it would allow for a best-fit model with lower $\dot{M}_2^h$ value and corresponding larger $M_1$ and $i$ values.

It is worth noting that the shape of the enhanced mass event is simple, set by one $\dot{M}_2$ value during quiescence and another in the high-accretion state; these values should correspond to an average $\dot{M}_2$ over the period. In the real case, the event is expected to show fluctuations and morphologies throughout the transient.

\section{Discussion} \label{sec:disc}

\subsection{Are there alternative explanations for the high-brightness state?} \label{sec:hbs}

Here we address two alternative explanations for the 15\,yr long high-brightness state of T~CrB, namely whether it could be the consequence of a dwarf-nova type outburst powered by the thermal-viscous disk instability or the result of a steady nuclear burning phase at the WD surface.

The disk-instability model (DIM) assumes that matter accumulates in a low-viscosity disk ($\alpha_c \sim 0.01$) during quiescence, with gas temperature and surface density $\Sigma$ progressively increasing everywhere, until a critical surface density is reached at some radius, bringing the disk into a higher viscosity ($\alpha_h \sim 0.1-0.2$) regime where the accumulated matter quickly accretes onto the WD \citep[e.g.,][]{Lasota2001}. The mass transfer rate, $\dot{M}_2$, remains constant during the whole outburst cycle. For the case of T~CrB, we may estimate $\dot{M}_2$ from the envelope mass that needs to accumulate onto the WD surface in order to trigger the next nova eruption and from the recurrent nova timescale,
\begin{equation}
\dot{M}_2(\mathrm{DIM})= \frac{M_{ig}}{T_R }=
 \frac{2.57\times 10^{-6} M_\odot}{80\,\mathrm{yr}} =
  3.2\times 10^{-8} M_\odot \, \mathrm{yr}^{-1}  \, .
\end{equation}
In this scenario, $\dot{M}_\mathrm{low}= 2\times 10^{-9} \, M_\odot$\,yr$^{-1}$ becomes the residual mass accretion onto the WD in quiescence. Accordingly, $\dot{M}_\mathrm{low} \ll \dot{M}_2$, and the effective rate at which matter accumulates in the disk is $\dot{M}_\mathrm{accum}= \dot{M}_2 - \dot{M}_\mathrm{low}= 3 \times 10^{-8} \, M_\odot$\,yr$^{-1}$. Over a quiescence interval of $T_q= (65\pm 3)$\,yr, the disk accumulates a mass of $\Delta M_\mathrm{disk}= (1.95 \pm 0.09)\times 10^{-6} M_\odot$. For an outburst mass accretion rate of $\dot{M}_\mathrm{out}= \dot{M}_2^h= 1.9\times 10^{-7} \, M_\odot$\,yr$^{-1}$, it takes $T_\mathrm{out}= (10.3\pm 0.5)$\,yr to dump the accumulated matter onto the WD. This predicted DIM outburst duration is significantly shorter than the overall length of the high-brightness state at the $5\sigma$ confidence level, indicating that not enough mass could be accumulated in the accretion disk during the quiescence period to sustain a DIM outburst for the observed duration.

On the other hand, the nova eruption occurs roughly half-way the high-brightness state phase, probably ejecting not only the WD envelope mass but also the whole accretion disk mass in the process. Given that the high-brightness state resumes after the end of the T~CrB nova eruption, this poses a double problem for the DIM scenario. First, it is hard to explain how a critical surface density condition which takes about 65\,yr to occur previous to the nova eruption can repeat itself right after the short nova eruption. Second, the maximum amount of disk mass that could accumulate after the disk disruption (assuming that the disk starts refilling just after nova maximum) would account for an outburst duration (at the $\dot{M}_\mathrm{out}$ rate) of $\leq 60$\,d, two orders of magnitude shorter than the observed 7\,yr length of the high-brightness state after the nova eruption. This makes the DIM scenario a non-viable explanation for the high-brightness state in T~CrB.

The alternative explanation of the high-brightness state as the consequence of a steady nuclear burning phase at the WD surface also has several problems. First, in order for steady nuclear burning to occur at the surface of a $M_1= 1.29\,M_\odot$ WD the mass accretion rate needs to be $\dot{M}_\mathrm{burn} \geq 2.3\times 10^{-7} \, M_\odot$\,yr$^{-1}$ (see Figure~\ref{fig:diagrama}), $\geq 20$\% larger than inferred from the modeling of the observations. The resulting amplitude of the high-brightness state as well as the magnitude of the irradiation effect would be systematically larger than observed. Second, the predicted luminosity from steady H-burning at the surface of a $M_1= 1.29\,M_\odot$ WD is $L_\mathrm{nuc} \simeq 24\times 10^3\, L_\odot$ \citep{Wolfetal2013}, a factor of $\simeq 30$ larger than the inferred luminosity of the accretion disk at the high-brightness state, $L_\mathrm{disk}= 8.3\times 10^2\,L_\odot$, implying that T~CrB should be $\simeq 3.7$\,mag brighter than observed in the blue. Third, the strong variability observed in soft X-ray observations during the high-brightness state as well as the inferred small area of the corresponding blackbody emitting region derived from spectral modeling indicates that this emission is not powered by H-burning at the WD surface \citep{Lunaet2018}. Fourth, there would be no explanation for why the mass accretion rate suddenly increases by a factor $\simeq 100$ from $\dot{M}_\mathrm{low}$ to $\dot{M}_\mathrm{burn}$ some 8\,yr before the nova eruption and why would it go back to $\dot{M}_\mathrm{low}$ about 7\,yr after the nova eruption (other than assuming it is an enhanced mass-transfer event from the RG). Finally, the transition between the steady and unsteady nuclear burning regimes is very sharp, making these two possibilities mutually excludent \citep{Paczynski1983,Wolfetal2013}. Steady nuclear burning consumes the hydrogen-rich envelope mass while removing the envelope degeneracy essential for the TNR, thereby preventing a nova eruption to occur while the system is in a steady H-burning phase -- whereas in T~CrB the nova eruption is superimposed on the high-brightness state. All these discrepancies make the steady nuclear burning scenario also a non-viable explanation for the high-brightness state in T~CrB.

\subsection{The high-brightness state and the nova eruption \label{sec:hmdot}} 

The basic hypothesis of our modeling is that the high-brightness state is a unique event which starts 8\,yr before and lasts until 7\,yr after the T~CrB nova eruption, and that the decline in brightness observed $\simeq 1$\,yr before the eruption \citep[the 'pre-eruption dip',][] {Schaeferetal2023} is part of the eruption itself (see next section).

During the nova event, the accretion disk is probably ejected together with the envelope accumulated onto the WD surface. At an expansion velocity of $\simeq 4000$\,km\,s$^{-1}$\citep{Sanford1946,McLaughlin1946}, the ejecta has expanded beyond the binary less than a day after the onset of the eruption. From this moment onwards, the RG is presumably able to restore mass transfer at the high-accretion rate, $\dot{M}_2^h$, and to restablish the accretion disk on a timescale of $\leq 400$\,d. This is a higher limit to the timescale for the system to return to the previous high-brightness state because irradiation of the RG by the hot and extended primary may temporarily increase the mass transfer rate beyond $\dot{M}_2^h$ and accelerate the restablishment of the accretion disk.

The characteristics of the T~CrB nova eruption \citep[e.g.,][]{schaefer2010} are reminiscent of those observed in DQ~Her \citep{McLaughlin1960,Warner1995} and FH~Ser \citep{Bode1982}, where the initial decline is interrupted by a transition phase (marked by a deep minimum of amplitude $\geq 4$\,mag and duration of 2-3 months), followed by a brightness increase and recovery of the previous decline rate (leading to a secondary maximum). In FH~Ser the brightness minimum coincides with an excess IR emission, indicating that the visual minimum is caused by the formation of absorbing dust grains which reradiate the intercepted energy in the IR while the bolometric luminosity remains constant \citep{Bode1982,Warner1995}. Similarly, in T~CrB the secondary maximum of 1946 coincides with the rise of a shell spectrum implying the formation of an optically thick envelope which absorbs and converts the UV radiations from the hot primary in increased visible flux \citep{selvelli92}. In DQ~Her and FH~Ser the ejected envelope shows an elongated geometry with an equatorial belt and polar blobs \citep{Mustel70,Hutchings1972}; optical imaging of T~CrB also show a bipolar structure \citep{Williams1977}. In addition, the cousin, fast recurrent nova RS~Oph shows a clear transition phase \citep[identified as a `post-eruption dip' by][]{schaefer2010} starting $\simeq 70$\,d and lasting until $\sim (450-500)$\,d after the onset of its eruption. Spectropolarimetric monitoring of its 2021 eruption \citep{Nikolovet2023} shows that dust grains roughly aligned with the binary orbital plane were present in RS~Oph more than 80\,d after the onset of the eruption (i.e., during its transition phase), indicating that dust grains also form in low-mass, high-velocity ejecta of recurrent novae similar to T~CrB. Indeed, a recent statistical study of dust formation in novae finds that 17 percent of very fast novae do form dust grains \citep{Chonget2025}. Having recurrent, fast nova eruptions with a transition phase caused by dust grain formation in 2 (RS~Oph and T~CrB) out of $\simeq 10$ known recurrent novae is consistent with their statistics.

Thus, the primary and the secondary observed maxima of T~CrB might not be two separate events demanding a challenging and unusual explanation \citep{Webbink1987}, but two phases of a unique nova eruption event. If this interpretation is correct, the decline phase ends about 400\,d after the start of the T~CrB nova eruption \citep[see Figure~21 of][] {schaefer2010}, when the accretion disk is already back to the high-brightness state (and the H-rich envelope for the next eruption starts to accumulate on the WD surface). In favor of this interpretation, we note that the minimum of the transition phase does not correspond to the high-brightness state; at this phase, T~CrB is fainter in the B-band and brighter in the V-band than during the high-brightness state that precedes the 1946 nova eruption, with colors $(B-V)_h \simeq +0.5$\,mag and $(B-V)_t \simeq +1.0$\,mag, respectively in the high-brightness state and during the transition phase \citep[see Figure~23 of][]{schaefer2010}.

An alternative interpretation of the observed secondary maximum of T~CrB is that it is caused by irradiation of the RG by a cooling, hot and extended WD \citep[but with constant radius,][]{Munari2023b}. In this scenario, the heating of the inner face of the RG (up to $\simeq 7800$\,K) will likely increase its mass transfer rate and help to shorten the timescale for the restablishment of the accretion disk, ensuring that T~CrB is back at the high-brightness state by the end of the overall nova decline phase.

\subsection{On the pre-eruption dip}

If there is a unique 15\,yr long event of enhanced mass accretion around the nova eruption, what explains the brightness decline over the $\simeq (1-2)$\,yr previous to the nova eruption? Here we put forward an explanation for the pre-eruption dip based on the observations.

Recent observations indicate that the pre-eruption brightness decline of T~CrB is much more pronounced in the blue than in the red \citep{munari23,Stoyanovetal2024}, leading \citet{Schaeferetal2023} to suggest that the brightness decline is the result of enhanced absorption by dust grains from a short and anticipated event of mass ejection. However, the lack of evidence for enhanced IR emission \citep{Woodwardetal2023} and the fact that the amplitude of the ellipsoidal modulation from the distorted RG remains constant along the brightness decline phase \citep{Schaeferetal2023} indicates that it is the accretion disk which is fading in brightness (by a factor of $\simeq 100$, in order to bring it back close to its quiescent brightness level).

There are two possible ways to decrease the accretion disk luminosity: one is by decreasing the mass-transfer rate and the other is by increasing the inner disk radius. A reduction in mass accretion rate by a factor $\simeq 100$ (from $1.9\times 10^{-7} \, M_\odot$\,yr$^{-1}$ back to $2\times 10^{-9} \, M_\odot$\,yr$^{-1}$) results in brightness variations of approximately the same magnitude across the $BVRI$ passbands -- in contradiction with the observations. On the other hand, an increase in inner disk radius by the same factor leads to a color-dependent brightness variation, more pronounced in the $B$-passband and of progressively lower amplitude with increasing wavelength -- in good agreement with the color changes observed during the pre-eruption dip. This is the consequence of the fact that increasing the inner disk radius selectively eliminates the hottest and bluest inner disk regions. Furthermore, increasing the inner disk radius by a factor $\simeq 100$ also increases the volume of a geometrically thin, equatorial boundary layer by a comparable amount, allowing it to become optically thin during the pre-eruption dip -- again in good agreement with the observations \citep{Stoyanovetal2024}. Thus, the observed brightness changes during the pre-eruption dip favors the interpretation that it is caused by an expanding inner disk radius.

In addition, assuming that the Doppler line broadening is caused by the Keplerian rotation of gas in the accretion disk, the half width at zero intensity (HWZI) of the emission lines provides a useful estimate of the inner disk radius and allow us to further test the idea that the pre-eruption dip is the consequence of an expanding inner disk radius. For example, the temperatures of the opaque steady disk at the high-brightness state are sufficiently low to allow the formation of hydrogen emission lines ($T \leq 10000$\,K) for $R \geq 1.06\,R_\odot$, corresponding to Keplerian velocities of $v_\phi \leq 485$\,km\,s$^{-1}$. Accordingly, the observed width of the Balmer lines during the high-brightness state is $\mathrm{HWZI} \leq 500$\,km\,s$^{-1}$  \citep{Munariet2016,Zamanovet2024}, in good agreement with this estimate. If the mass-transfer rate is decreased by a factor $\simeq 100$, the systematic reduction of accretion disk temperatures at all radii allows the extension of the hydrogen line emitting region down to $R \geq 0.22\,R_\odot$, corresponding to significantly higher Keplerian velocities, $v_\phi \leq 1050$\,km\,s$^{-1}$. However, spectroscopy at the minimum of the pre-eruption dip \citep[at 2023 October,][]{Zamanovet2024} shows that the width of the H$\alpha$ line has instead {\em reduced} to $\mathrm{HWZI} \simeq (230-250)$\,km\,s$^{-1}$, corresponding to an inner disk radius for hydrogen emission of $R_{in}\simeq (3.9-4.6)\,R_\odot$, a fator $\simeq 20$ times larger than predicted by the decreasing mass-transfer rate scenario. This suggests a significant increase in inner disk radius from the high-brightness state to the pre-eruption dip. Assuming that the inner disk radius at the high-brightness state equals the WD radius and taking $R_{in}$ as an upper limit to the inner disk radius at 2023 October, we infer an upper limit to the average expansion velocity of $v_\mathrm{exp} < (0.13-0.16)$\,km\,s$^{-1}$ along the $\simeq 8$\,months from the start to the minimum of the pre-eruption dip \citep[from 2023 February to 2023 October,][]{munari23,Zamanovet2024}.

\begin{figure}[!ht]
\centering
\includegraphics[width=1.0\columnwidth]{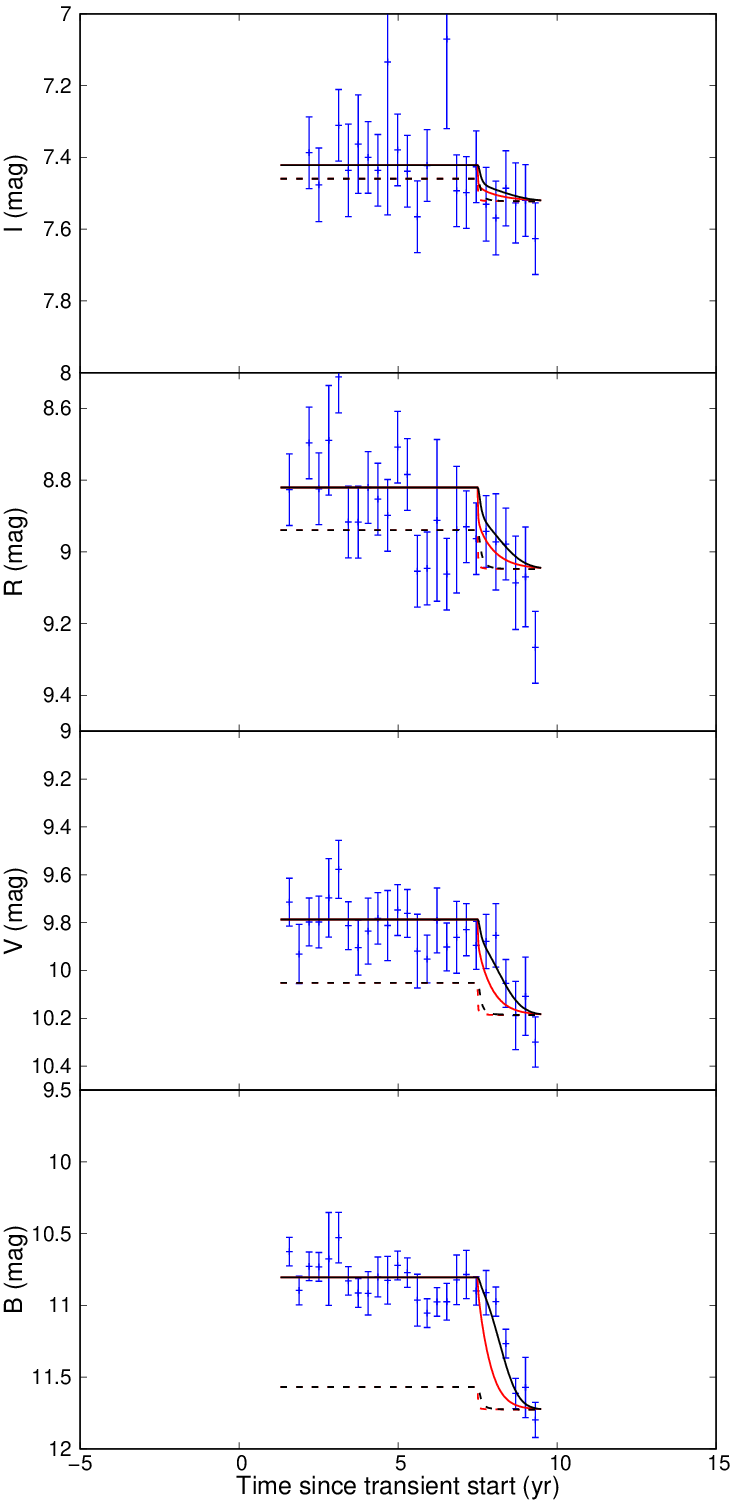}
\caption{Simulation results with expansion of the inner radius with constant acceleration (black solid curve) and constant velocity (red solid curve). The points with error bars are the same as in Figure~\ref{fig:dados} for $B$ band and analogous for the other bands. The black and red dashed curves are the contribution from the irradiated red giant companion in the models with constant acceleration and constant velocity, respectively. \label{fig:simulacao_interno}}
\end{figure}

We further tested the idea of such a slowly expanding inner disk radius by simulating the evolution of the accretion disk of model\,II with a variable inner radius, at a fixed mass-transfer rate $\dot{M}^h_2$. Two cases were considered. In the first case, the inner disk radius increases from $R_{in}= R_{WD}= 0.0045\,R_\odot$ up to a few $R_\odot$ at constant acceleration, corresponding to an average expansion velocity $v_\mathrm{exp}$ over a 2\,yr time interval. In the second case, the inner disk radius expands at a constant velocity $v_\mathrm{exp}$ during the same time interval. The results are compared to the observations of the pre-eruption dip in Figure~\ref{fig:simulacao_interno}. The observed brightness decline in the $BVRI$ passbands is reasonably well reproduced by the model with accelerated inner disk expansion. A small average expansion velocity of $v_\mathrm{exp}= 0.02$\,km\,s$^{-1}$ provides the best-fit to the observational data, in agreement with the upper limit estimated from the observations.

The above observational picture indicates that the decline in accretion disk luminosity associated to the pre-eruption dip is not caused by a decrease in mass-transfer rate but by a significant, slow increase in inner disk radius during this time. What could cause the slow inner disk radius expansion?

The sequence of events that leads to a nova eruption can be described by three phases \citep[e.g.,][]{Prialnik1986,ShenBildsten2009,Wolfetal2013}. The pre-nova {\em accretion phase} ends when the accumulated envelope mass reaches the ignition conditions. The {\em convection phase} starts with the onset of nuclear reactions at the base of the accreted envelope. Since the gas is degenerate, the H-burning region spreads rapidly both inwards and outwards, the nuclear luminosity $L_\mathrm{nuc}$ rises exponentially, temperature gradients increase and the H-burning region becomes convectively unstable. The timescale for the development of convection throughout the accreted envelope (i.e., the duration of the convection phase) is short compared to the thermal timescale. Therefore, the envelope is out of thermal equilibrium; its bolometric luminosity remains almost constant ($L_\mathrm{bol}\sim 10^{-2}\,L_\odot$) while $L_\mathrm{nuc}$ increasingly exceeds $L_\mathrm{bol}$ along this phase \citep[e.g.,][]{Prialnik1986}. The end of the convection phase sets the start of the {\em explosion phase}, with the TNR on a fully convective envelope leading to its fast expansion and significant increase in luminosity identified as the nova eruption. The duration of the convection phase is 20-100 times shorter than that of the preceding accretion phase \citep{Prialnik1986,ShenBildsten2009}. For T~CrB, the accretion phase of which lasts about 80\,yr, this leads to an estimated length of $(0.8-4.0)$\,yr, in good agreement with the observed $\simeq (1-2)$\,yr duration of the pre-eruption dip. Hence, {\em the pre-eruption dip occurs during the convection phase in T~CrB}.

We suggest that the inferred expansion of the inner disk radius is connected to the lack of thermal equilibrium and the development of strong convection in the accreted envelope during this phase, in two possible scenarios. We first note that, at the onset of the convection phase on top of a high-mass WD fed at high accretion rates, the gas is partially degenerate even at the base of the envelope \citep[$T_b\simeq 3\times 10^7$\,K and $\rho_b\simeq 10^3$\,g\,cm$^3$, e.g.,][]{ShenBildsten2009} and gas pressure is sensitive to temperature. According to the first law of thermodynamics, the additional heat provided to each layer across the accreted envelope drives changes in their temperature and density at rates \citep{KippenhahnWeigert1990},
\begin{equation}
\frac{dq}{dt} = c_\rho \frac{dT}{dt} + c_T \frac{d\rho}{dt} \, ,
\end{equation}
where
\begin{eqnarray}
c_\rho & = & \left( \frac{\partial u}{\partial T} \right)_\rho > 0 \\
c_T   & = & \left( \frac{\partial u}{\partial\rho}\right)_T
        - \frac{P}{\rho} < 0 \, ,
\end{eqnarray}
$c_\rho$ and $c_T$ are, respectively, the specific heats at constant density and temperature, $u$ is the internal energy per mass and $P$ is the total pressure. A $dq/dt>0$ implies $dT/dt>0$ (because $c_\rho>0$) and $d\rho/dt<0$ (because $c_T<0$). Thus, the first possible scenario is that the increasing excess heat deposited at the accreted envelope during the convection phase continuously increases temperature (leading to pressure unbalance, loss of hydrostatic equilibrium, and expansion) and decreases density (i.e., direct expansion), driving the envelope into slow (but accelerated) expansion at roughly constant bolometric luminosity while convection spreads and efficiently transports excess nuclear energy further upwards
\footnote{This is also true in the case of isobaric expansion, where the increase in temperature is compensated by a matching decrease in density (meaning envelope expansion) in order to keep pressure constant.}.
If the inner disk radius coincides with the surface of the accreted envelope, the expansion of the envelope leads to the progressive increase of the inner disk radius
\footnote{The volumetric densities at the inner radius of a steady opaque disk with $\alpha=3$ and $\dot{M}_2^h$ are $\rho_d < 10^{-9}$\,g\,cm$^{-3}$, at least two orders of magnitude lower than those at the surface of a B0 main-sequence star of $R=6 \, R_\odot$ \citep[e.g.,][]{Vitense}, indicating that while the expanding atmosphere of the primary star has $R\leq 20\,R_\odot$ it will be dense enough to prevent the inner disk regions to lag behind and extend inwards of the expanding atmosphere.}
and to the continuous decrease of the accretion disk luminosity even if the mass accretion rate remains constant during the process.

If this scenario is correct, why there is no report of envelope expansion during the convection phase in nova eruption simulations? Numerical simulations of nova eruptions are computationally very demanding \citep[e.g.,][]{Jose2020} and, for computational feasibility, most numerical simulations of nova eruptions have been performed with 1D codes, with two relevant limitations. First, hydrostatic equilibrium is imposed throughout the accreted envelope during the convection phase, until the relative pressure unbalance exceeds a threshold (usually $dP/P > 10^{-5}$) and the simulation switches to full hydrodynamic calculations in order to follow the explosion phase \citep[e.g.,] [and references therein]{Prialnik1986,Glasner1997,Zingale2002,Jose2020}. By design, these simulations are unable to identify and to account for a slow accretion envelope expansion during the convection phase. While this choice is computationally convenient and justified by the focus on the ensuing TNR, it leads to the unrealistic behavior of an accretion envelope which remains static for most of the convective phase and suddenly starts expanding at already significant velocities $v_\mathrm{exp}\simeq (10^1-10^2)$\,km\,s$^{-1}$. Signs that imposing hydrostatic equilibrium throughout the convection phase leads to an incomplete physical description and that envelope expansion do occur appear when 1D nova eruption simulations are used as starting point to 2D full hydrodynamic simulations: \citet{Casanova2018} reported that forcing the initial 1D model to be in hydrostatic equilibrium after mapping into 2D imposes a restriction on the portion of the envelope that can be computed, since the density dramatically decreases and produces numerical underflow \citep[see also][]{Zingale2002} -- suggesting that the 1D model envelope is undersized for its physical conditions (i.e., there is unaccounted expansion in the 1D model). 1D modeling is further limited by the use of the mixing-length theory of convection (MLT), which cannot account for either mass flux \citep[i.e., envelope expansion,][]{Jose2020} or strong mixing at core-envelope interface \citep{Casanova2011}. 3D simulations suggest that strong turbulence and mixing do occur at the core-envelope interface and that the resulting convective velocities are systematically larger (by factors 2-3) than those obtained assuming MLT \citep{Jose2020}, driving significant dredge-up of CO or ONeMg material from the WD core into the H-burning envelope (which increases the rate of nuclear energy generation) and efficient transport of excess nuclear energy to upper, less degenerate envelope layers -- both of which may contribute to envelope expansion during the convection phase.

An alternative scenario to explain the inferred increase of the inner disk radius is related to the possible development of a strong, transient magnetic field during the convection phase. \citet{Casanova2016} suggested that there may be sufficient time during the vigorous convection phase to build a fast dynamo and a strong magnetic field, which could then be advected together with the ejecta during the ensuing nova eruption. If this is the case, the magnetospheric radius increases with magnetic field strenght along the convection phase, disrupting the accretion disk at progressively larger radii and leading to the continuous decrease of the accretion disk luminosity even if the mass accretion rate remains constant during the process. For a magnetic WD, the magnetospheric radius inside which magnetic pressure exceeds ram pressure and the accretion disk is disrupted is given by \citep{ACPower},
\begin{equation}
R_M \simeq 5.1\times 10^8 \,\dot{M}^{-2/7}_{16}
\left( M_1/M_\odot \right)^{-1/7} \mu^{4/7}_{30} \,\mathrm{cm} \, ,
\end{equation}
where $\dot{M}_{16}$ is the mass-accretion rate in units of $10^{16}$\,g\,s$^{-1}$, $\mu_{30}$ is the magnetic moment ($= B_{WD} R_{WD}^3$) in units of $10^{30}$\,G\,cm$^3$, and $B_{WD}$ is the WD surface magnetic field. For the high-brightness state of T~CrB, $\dot{M}_{16}= 1200$ and $R_M = 1.73\times 10^5\,B_{WD}^{4/7}$\,cm. Thus, in order for $R_M$ to increase from $\leq R_{WD}$ up to $\simeq 100\,R_{WD}$, $B_{WD}$ needs to increase from $\leq 5 \times 10^5$\,G up to $1.6\times 10^9$\,G. The larger field corresponds to a magnetic moment of $\mu \simeq 5\times 10^{34}$\,G\,cm$^3$, comparable to those of strong field polars \citep[e.g.,][]{Warner1995}. The corresponding fundamental frequency for cyclotron emission is $f_{cyc}= 4.5\times 10^{15}$\,Hz ($\lambda_{cyc}=66.7$\,nm). The predictions from this scenario can be observationally tested: enhanced EUV and soft X-ray emission with cyclotron humps (in the range $19-190$\,eV for the first 10 cyclotron harmonics), linear and circular polarization (and possibly X-ray variability at the WD rotation period) associated to the pre-eruption dip in T~CrB. However, none of these has been detected during the current dip. Soft X-rays were not detected, while near UV emission fainted (Luna et al., {\em in prep.}) and intrinsic optical polarization was absent (Y. Nikolov, {\em priv. comm.}). We thus do not favor this scenario. 

If the pre-eruption dip is a general consequence of the convection phase, why there is no evidence of pre-eruption dips on all nova? What is special about T~CrB?

T~CrB is the only (recurrent) nova whose eruption starts while the system is in a state of enhanced mass accretion, with its disk luminosity being about 100 times brighter than in quiescence. Perhaps the lack of detection of pre-eruption dips in novae is the effect of an observational bias which will be reduced with the new all-sky surveys. For example, given that all other recurrent nova go into eruption while their accretion disks are at a much lower brightness level, their pre-eruption dip may easily be diluted in the brightness of the RG companion and in the typical $\simeq 1$\,mag brightness variations that affect their light curves \citep{schaefer2010}. Moreover, with accretion rates of $\simeq 10^{-10}-10^{-9}\,M_\odot$\,yr$^{-1}$, small accretion disks and sub-solar companions, quiescent classical novae are intrinsically fainter than T~CrB and beyond the reach of visual observations and of most amateur astronomers, which has prevented well sampled photometric monitoring of their pre-eruption stage. Furthermore, because of the much longer accretion phase, their convection phase is expected to last about 1-2 centuries; there is no comprehensive and precise photometry of pre-eruption novae over this large amount of time. Nevertheless, \citet{Zamanovet2024b} has reported the tantalizing detection of a pre-eruption dip in RS~Oph (although the authors arrived at a different interpretation about its origin than the one we put forward here). Finally, detailed optical observations during the inter-eruption phases of a recurrent nova with a short recurrence period, and thus with a short convection phase, such as RN~M31N~2008-12a \citep{Darnley2014} would provide a clear-cut test for the scenario proposed above.

\subsection{Which came first, the chicken or the egg?}

\citet{Schaefer2023_mnras} raised the interesting issue that ``The existence of the pre-eruption high-state presents a mystery as to why the system brightens before the eruption", and then concluded that ``So we are left with no explanation for the enigma as to why the high-state anticipates the eruption by 10\,yr, just as we have no explanation for the cause or physical mechanism of the high-state existence". This apparent enigma is a consequence of an inversion of the relevance of the events in T~CrB.

As remarked in Section~\ref{sec:int}, without a phase of enhanced mass-accretion, T~CrB would be a classical nova with a much longer interval between successive eruptions ($\simeq 5480$\,yr). The best explanation for the high-brightness state (Section~\ref{sec:hbs}) is that it corresponds to the crucial phase of enhanced mass-accretion which determines the much shorter recurrence interval of the T~CrB nova eruptions. It is thus correct to say that the TNRs of T~CrB are powered by accretion events. Indeed, the matter accumulated onto the WD surface during the 15\,yr long enhanced accretion phase responds for 95\% of $M_{ig}$, indicating that the probability that the ignition conditions be reached (and lead to the nova eruption) during the high-brightness state is also of 95\%. This provides a natural explanation for why the nova eruption is superimposed on the high-brightness state. In other words, it is the increased mass accretion phase during the high-brightness state (the cause) that drives the nova eruption (the consequence), underscoring previous suggestions by \citet{Lunaet2020} and \citet{Zamanovet2023}, and solving the apparent enigma raised by \citet{Schaefer2023_mnras}.

Concerning the recurrence of the enhanced mass-accretion events, a possible explanation for the sudden increases in $\dot{M}_2$ required by MTIM involves the interaction of starspots with the inner lagrangian point $L_1$. While the passage of starspots in front of the $L_1$ point can significantly reduce $\dot{M}_2$
\citep{LivioPringle1994,KingCannizzo1998}, events of enhanced  $\dot{M}_2$ may occur when there are no starspots transiting $L_1$ \citep{HameuryLasota2014}.

A possible mechanism through which the high-accretion state develops and repeats itself every $\sim$80\,yr is as follows. A starspot-free region on the secondary that periodically pass in front of L$_1$ as seeing from the WD would enhance the accretion through L$_1$. Lets assume that P$_{spot}$ is the period of the starspot-free region which is slightly larger than the orbital period P$_{orb}$, then the recurrence time can be determined from the beat period between both, 1/P$_{rec}$= 1/P$_{orb}$ - 1/P$_{spot}$. Defining $\epsilon$=(P$_{spot}$-P$_{orb}$)/P$_{orb}$ as the relative difference between the orbital period and the period of the starspot-free region, we find that $\epsilon$=P$_{orb}$/(P$_{rec}$-P$_{orb}$). Although of different spectral type than the secondary in \tcrb\, there is evidence of starspots on the secondary of SS~Cyg, a well-known dwarf nova \citep{webb2002}. In the case of SS~Cyg, where P$_{orb}$= 6.6\,h and the dwarf nova outburst recurs on a characteristic timescale of P$_{rec} \simeq 40$\,d $\simeq 960$\,h, we find $\epsilon$= 0.007. In the case of T~CrB, with P$_{orb}$= 227.5\,d, P$_{rec}= 80$\,yr $= 29220$\,d, we find $\epsilon$= 0.008. This suggests that, while in the case of SS~Cyg a shift of 0.7\% between the starspot and orbital periods can explain its dwarf nova outburst recurrence timescale, in the case of \tcrb\ a comparable shift could explain the recurrence time of the high-accretion state. Although appealing, this mechanism calls for further observations to be accepted, but offers a path to explain the recurrence of the high-accretion state during which the TNR happens. A search for starspot-free regions on the surface of the RG, their evolution with time and possible correlation with the high-brightness state may help to elucidate if the proposed scenario is viable.

\section{Conclusions} \label{sec:concl}

The observed T~CrB brightness variations during the high-accretion state can be satisfactorily reproduced by models of an enhanced mass-transfer event of duration $\Delta t_e = 15$\,yr onto a high-viscosity accretion disk with $\alpha = 3$. The best-fit, self-consistent model has an inferred WD mass $M_1 = 1.29\, M_\odot$, an inclination $i = 57.3^\circ$, and quiescent and high-accretion mass-transfer rates, respectively, of $2.0\times 10^{-9} \, M_\odot$\,yr$^{-1}$ and $1.9 \times 10^{-7} \, M_\odot$\,yr$^{-1}$. This model also provides a good description of the colour variations of T~CrB throughout the transient.

The pre-eruption dip occurs during the convection phase that precedes the TNR in T~CrB and is best described by a slow, accelerated expansion of the inner disk radius at an average velocity of 0.02\,km\,s$^{-1}$ over a 2\,yr timescale. We suggest that this expansion is connected to the lack of thermal equilibrium and the development of strong convection in the accreted envelope during this phase, in two possible scenarios: (1) the slow expansion of the accreted envelope itself (pushing the inner disk radius progressively outwards) and (2) the progressive disruption of the inner disk regions by the magnetosphere of a transient magnetic field. However, this last scenario does not seem to be favored by the current observational data.

\begin{acknowledgments}
We thank an anonymous referee for interesting comments and suggestions which led to the expansion of the scope of the paper and to an improved presentation of our results.
This study was financed in part by the Coordenação de Aperfeiçoamento de Pessoal de Nível Superior - Brasil (CAPES) - Finance Code 001. WS acknowledges financial support from CNPq/Brazil (Proc. 300834/2023-3, 301366/2023-3, 300252/2024-2 and 301472/2024-6). GJML is member of the CIC-CONICET (Argentina).
\end{acknowledgments}

%

\vspace{5mm}
\facilities{AAVSO}




\bibliography{ref}{}
\bibliographystyle{aasjournal}



\end{document}